# Current spinon-holon description of the one-dimensional charge-transfer insulator $SrCuO_2$: Angle-resolved photoemission measurements

A. Koitzsch[1,2], S. V. Borisenko[1], J. Geck[1], V. B. Zabolotnyy[1], M. Knupfer[1], J. Fink[1], P. Ribeiro[1], B. Büchner[1,2], R. Follath[3]

[1] *IFW Dresden, P.O.Box 270116, D-01171Dresden, Germany*

[2] *Institute for Solid State Physics, TU Dresden, D-01062 Dresden, Germany*

[3] *BESSY GmbH, Albert-Einstein-Strasse 15, 12489 Berlin, Germany*

**Abstract**: We have investigated the low-energy electronic structure of the strongly correlated one-dimensional copper oxide chain compound $SrCuO_2$ by angle resolved photoemission as a function of excitation energy. In addition to the prominent spinon-holon continuum we observe a peaklike and dispersive feature at the zone boundary. By fine-tuning the experimental parameters we are able to monitor the full holon branch and to directly measure the electronic hopping parameter with unprecedented accuracy.

One-dimensional cuprates attract the attention of solid-state physicists for many reasons. Among them almost perfect examples of one dimensional spin ½ Heisenberg chains can be found [1, 2]. It is well known that the standard models for the description of their low-energy electronic excitations spectacularly fail: a single electronic excitation decomposes into two independent excitations carrying either spin (spinons) or charge (holons) [3, 4]. While being conceptually highly important in its own right, the spin-charge separation promotes another striking peculiarity of these systems, namely the large non-linear optical response, which renders these materials even technologically interesting [5, 6]. More specifically, external electrical fields tune the optical parameters enabling the construction of optical switches [7]. Furthermore, the close structural and electronic relationship of these materials to the high- temperature superconductors (HTSC), and the known instabilities of the HTSC towards one-dimensional phenomena, forms a background, which provides a strong motivation for close scrutiny [8].

The first direct observation of the spin-charge separation was achieved by angle-resolved photoemission (ARPES) in $SrCuO_2$ [9]. It was found that the bandwidth of the lowest electronic excitation in $SrCuO_2$ scales with $t$ , the electronic hopping parameter, rather than $J$, the exchange constant, in striking contrast to the two dimensional insulator $Sr_2CuO_2Cl_2$. This is a clear indication that the electronic excitations are decoupled from the spin background in the one dimensional case. Subsequent photoemission studies confirmed and substantiated the picture of the spin-charge separation [10, 11]. In particular, the spinon band has been explicitly resolved for the closely related compound $Sr_2CuO_3$ [12, 13]. Further optical [14], electron energy loss spectroscopy [15], inelastic neutron scattering studies [16], and inelastic x-ray scattering [17] are consistently interpreted within the framework of spin-charge separation.

However, major questions have been left open. From the very beginning spectral weight has been observed at the Brillouin zone boundary (ZB), which has been attributed to photoemission background [9], surface degradation [11] or the incoherent part of the spectral function in a one-band Hubbard model [18]. As will be shown below on the basis of our new results we can exclude these explanations, and propose instead deeper lying reasons for the ZB feature.

From the theoretical side the derivation of the spectral function of one-dimensional Mott systems has been the subject of intense effort [19, 20]. A crucial precondition for comparison with experiment is the reliable knowledge of the basic electronic parameters like $t$ and $J$. In this letter we show that previous low-energy photoemission studies tend to underestimate the holon bandwidth – and therefore $t$ – and provide a measurement of $t$ with unprecedented accuracy.

Eventually, the validity of the spin-charge separation description has been challenged recently by x-ray ARPES [18]. We explicitly show here that major predictions concerning the spinon and holon dispersion relations are fulfilled, which confirms the legitimacy of an explanation of the data within the framework of decoupled spin and charge degrees of freedom, although an extension of the theoretical models to account for the ZB feature will be necessary.

We report ARPES measurements on $SrCuO_2$, a one-dimensional charge transfer insulator whose copper-oxygen network consists of two weakly coupled corner-sharing chains. We investigated the low-lying electronic excitations by using a variety of excitation energies.

$SrCuO_2$ single crystals were grown using the travelling solvent zone technique in an optical furnace. A detailed description of the method can be found elsewhere [21].

The ARPES experiments were carried out using radiation from the U125/1-PGM beam line and an angle multiplexing photoemission spectrometer (SCIENTA SES 100) at the BESSY Synchrotron Radiation Facility and using a lab-based system equipped with an SES 200 analyzer and a Gammadata He discharge lamp. The photon energy for the synchrotron experiments ranged from hν = 90 to 100 eV with an energy resolution between 30 and 40 meV and an angular resolution of 0.2°. Resolution of the lab based experiments with hν = 21.2 eV and 40.8 eV was set to approx.. 30 meV with an angular resolution of 0.3°. $SrCuO_2$ single crystals were cleaved *in situ* at room temperature resulting in a mirror like cleavage surfaces. LEED (Low energy electron diffraction) pictures show sharp spots reflecting the orthorhombic crystal structure. As $SrCuO_2$ is an insulator the samples tend to charge up under photon flux. For some of the presented data we accepted small energy shifts in order to maximize the count rate. It was always confirmed that no charge-induced deformation of the line shape was present.

In Fig. 1a the results of photoemission measurements with hν = 100 eV and the momentum parallel to the chain direction are presented in a false color plot. The polarization of the incoming radiation was perpendicular to the chains and the spectra have been normalized for each *k* to the valence band [22]. The subject of our interest is the steeply dispersing, V-shaped structure, which is symmetric around the Γ-point. In accordance with previous studies [9] we identify this feature as the holon band [23]. It reaches its maximum energy around π/2 and disperses back in the second half of the BZ, where it rapidly looses spectral weight. The spinon band manifests itself as blurred intensity in the middle of the “V”. The significance of this dataset is twofold. First, it lays in the fact, that the holon band can be explicitly traced to the Γ-point, enabling observation of the full holon bandwidth. The red (dashed) line represents the theoretically expected holon dispersion according to $E = 2t \sin |k|$ (1) [19] with t = 0.71 eV. The same function has been overlaid to the low excitation energy (hν = 21.2 eV) ARPES data in Fig. 1b. It is clear, that a large part of the holon band is hidden below the valence band states in this case. This is due to the fact that for low excitation energies emission from the valence band, which is known to be primarily of O 2p character, is enhanced due to the larger crossection of O 2p orbitals for low-energy photons compared to high-energy photons [24]. Therefore, high photon energies are favorable to detect the full holon bandwidth.

Whereas the intensity of the holon band decreases in the second half of the Brillouin zone for the hν = 100 eV data this is not the case for the hν = 21.2 eV dataset, where the holon band is well monitored throughout the zone [25]. From this the good agreement between the dispersion relation (1) and the data becomes apparent, lending strong support to the underlying theoretical description within the framework of the spin-charge separation.

The second important feature of the dataset in Fig. 1a is constituted by the small but well resolved intensity at E ≈ 94.2 eV at the ZB. As is already clear from Fig. 1a, and will be discussed in detail below (see Fig.3), this feature is a well defined peak rather then a step like edge and it shows a finite dispersion, which excludes the possibility of an extrinsic origin.

Fig. 1a contains also the spinon dispersion (blue dashed-dotted line) with the functional form $E = -\pi J/2 \cos k$ [19] and J = 0.31 ± 0.08 eV. Compared to the determination of *t* the determination of *J* is more arguable because the spinon band manifests itself as a rela-tively broad shoulder as shown in the single curve representation of Fig. 1c. As can be observed in this figure, the two holon branches rapidly disperse towards the BZ center. For the green (labeled (1)) EDC only the holon peak is visible at E = 94.1 eV. For the blue (2) EDC a shoulder has developed at E ≈ 93.95 eV, which can be assigned to the spinon band, and eventually, at the zone center the holon band has dispersed away and merged with the main valence band, while the spinon band remains as a separate shoulder (red (3) EDC).

To analyze the reliability of the obtained value of the important hopping parameter $t$ let us consider Fig. 2. Here "band structure" representations of the low-lying electronic structure are depicted, obtained by taking the second energy derivative after moderate smoothing [31] of the dataset presented in Fig. 1a and, additionally, of an equivalent dataset taken with hν = 90 eV. Clearly a cusp-like feature at the Γ- point is observed in accordance with (1) supporting again the underlying description within the framework of the spin-charge separation. It is this cusp that defines the lower boundary of the holon branch. The upper boundary is given by the highest EDC peak maxima in the vicinity of (π/2), which is overlaid over the band structure map in Fig. 2 on the energy axes. $t$ results from half of the distance between the cusp and the white dashed line. From Fig. 2 and similar measurements for more samples we obtain $t = 0.71 \pm 0.04$ eV [30]. In Fig. 2 also the spinon band can be clearly seen as a broad ribbon in the middle of the holon "V".

Recently it has been proposed that for undoped two-dimensional cuprates the photoemission lineshape does not represent the "quasiparticle" but rather polaronic satellites [26, 27, 28, 29]. In principle such a mechanism could also apply for the one-dimensional cuprates. The lowest EDC in Fig. 1c has been fitted with a simple Gaussian lineshape, which results in good agreement. The equivalent Lorentzian fit is much worse (not shown). This is the same result as for the two-dimensional cuprates [26] where it has been taken as an argument in favor of the polaronic picture since the spectral function in the conventional self-energy description should rather be Lorentzian like. However, even in a polaronic picture the gross scheme of the data interpretation within the framework of the spin-charge separation would still be valid. If no spin-charge separation were present in the system, the bandwidth would be scaled by *J* as it is the case for the two-dimensional insulators, e.g. $Sr_2CuO_2Cl_2$ . Polarons cannot increase the bandwidth. The large bandwidth of the V-shape structure which scales with *t* rather then with *J*, is naturally explained by the holon-type dispersion. Nevertheless, as for the two-dimensional cuprates the polaronic scenario may account for the broad lineshapes observed.

One still has to keep in mind, however, that the lowest energy excitations for $SrCuO_2$ belong to a continuum. This is a complete different situation than for the two-dimensional cuprates. It is also possible that the existence of the continuum influences the lineshape. Further doping dependent studies are necessary to clarify the issue. With doping polaronic effects should become weaker and coherent excitations should emerge.

Now we turn to the detailed discussion of the ZB feature. Fig. 3a presents a blow-up of the corresponding region of Fig. 1a. The ZB feature forms a flat band around k=-1 which acquires a downward dispersion between k = -0.85 and k = -0.65. It is clearly distinct from the holon branch and its maximum kinetic energy lays approximately 140 meV below the top of the holon branch. Fig. 3b gives an EDC representation of the section of Fig. 3a between the vertical black lines. At k = -0.85  we find a peak at E = 94.3 eV. At k = -0.715 it has dispersed downwards by approx. 140 meV and merges with the holon band, which disperses upwards coming from lower energies. Fig. 3d compares EDCs from the ZB and the zone center indicating that the energy position of the ZB states is different from the spinon edge. The ZB feature can be also observed for lower excitation energy as is exemplified in Fig. 3c for hν = 40.8 eV. However, due to the increasing contribution from the valence band its relative intensity decreases.

Previously, a step-like feature has been reported around the ZB [9, 11] that has been assigned to photoemission background and surface degradation. On the basis of our data we exclude these possibilities. First, the feature is a well-defined peak and not just a step like background. Second, it shows dispersion and third a photoemission background should be largely k-independent but, as can be seen in Fig. 3d, we do not observe it at the Γ-point. Another possible origin of the ZB feature would be a surface state. However, recent high-energy angle-resolved photoemission [18] shows this feature even more pronounced. This points to its bulk nature, since high-energy ARPES has the advantage of an increased electron escape depth.

Having exhausted extrinsic explanations we are left with the option that the ZB feature is an essential part of the spectral function of this one-dimensional system, which has been missed in all previous theoretical derivations. From the basic electronic structure we do not expect bands other than the Cu $3d_{x^2-y^2}$/O $2p_{x,y}$ derived main "conduction" band, which forms the holon band, in this energy region. Therefore we are forced to search for the origin of the ZB feature within the spinon/holon description. During the photoemission process the photohole decays into a spinon and a holon. While the charge channel is protected by the large Mott gap, spin excitations with arbitrarily small energy remain possible. We speculate that the effect of spinon scattering, which allows in principle large momentum transfer, may be capable of transferring spectral weight to the ZB. Furthermore the presence of polaronic effects may qualitatively alter the distribution of spectral weight.

In conclusion we performed a photon energy dependent ARPES study of the one-dimensional charge transfer insulator $SrCuO_2$. Using excitation energy of hν = 100 eV the full holon branch can be monitored and the hopping parameter *t* can be estimated with high accuracy. Our results support the picture of a complete spin-charge separation. We observed and scrutinized an additional peak in the vicinity of the Brillouin zone boundary. In contrast to previous studies this feature has the character of a distinct peak with a finite dispersion. We argue that a consistent interpretation of this feature in terms of a photoemission background or a surface state cannot be achieved. Rather, we propose that it is a part of the spectral function, for the description of which the effects of spinon-scattering and the electron-phonon coupling could be essential and, possibly, also multi-band models are needed. We believe our experimental observation will be the basis of more complete theoretical concepts for one-dimensional cuprates.

We acknowledge financial support by the BMBF project 05KS4OD2/8 and the DFG under Grants SFB 463 and No. Kn 393/4 as part of the Forschergruppe FOR538. We thank R. Hayn, S.L. Drechsler and R. Kuzian for fruitful discussions.

[31] We applied a binomial smoothing procedure with a FWHM of 12 points.

## Captions

Fig. 1: (Color online) a) ARPES intensity plot as a function of wavenumber (in units of the Brillouin zone) and kinetic energy taken at photon energy hν = 100 eV at room temperature. The prominent V – shaped feature symmetric to the zone center represents the holon band. The red (dashed) and blue (dashed-dotted) lines correspond to the theoretical holon and spinon dispersion. There exists also weak intensity between 94 and 94.5 eV at the zone boundary. b) Similar data taken at hν = 21.2 eV. The holon band is depicted with identical parameters as in (a). Note that half of the bandwidth is hidden by the valence band for small excitation energies. The spinon band is not discernible. c) A section of the *k*-region shown in panel (a) presented as energy distribution curves (EDC). The green (labeled (1)) EDC shows the holon peak alone, the blue (2) EDC holon and spinon overlaid and the red (3) EDC the spinon alone. Part of the bottom EDC is fitted by a Gaussian.

Fig. 2: (Color online) a) "Band structure" presentation of the data shown in Fig. 1(a), obtained by the second energy derivative. Note the sharp cusp of the "V" confirming the description of the spectral function within the framework of spin-charge separation. The EDC with the peak position at the highest kinetic energy is overlaid on the energy axes.

The hopping parameter *t* is given by half of the distance between this peak and the cusp at the Γ-point. b) Similar dataset as in (a) but taken with hν = 90 eV.

Fig. 3: (Color online) a) Blow-up of the Brillouin zone boundary region of Fig. 1a showing a pronounced structure with clear dispersion between the vertical black lines. b) EDC representation of the marked region in (a). Blue circles highlight the peak maxima belonging to the holon band. Black dots highlight the ZB feature, which shows a small but finite disperion. c) ZB feature for data taken with hν = 40.8 eV. The ZB feature is still detectable but more difficult to observe due to the increased VB emission and due to the increased holon emission. d) Comparison of an EDC from the BZ boundary and the BZ center. The spinon shoulder appears at lower energy than the BZ boundary feature.

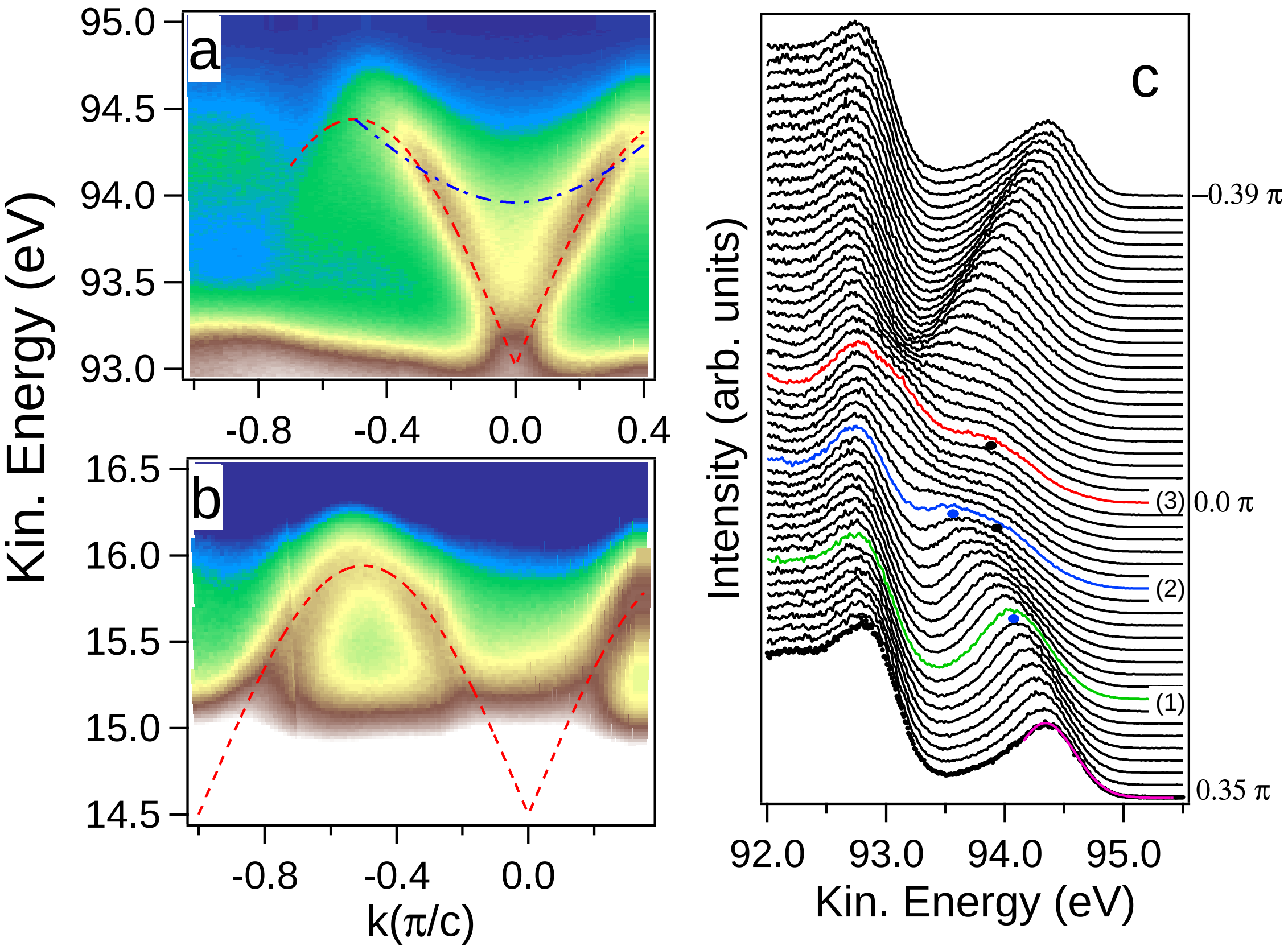

Fig. 1

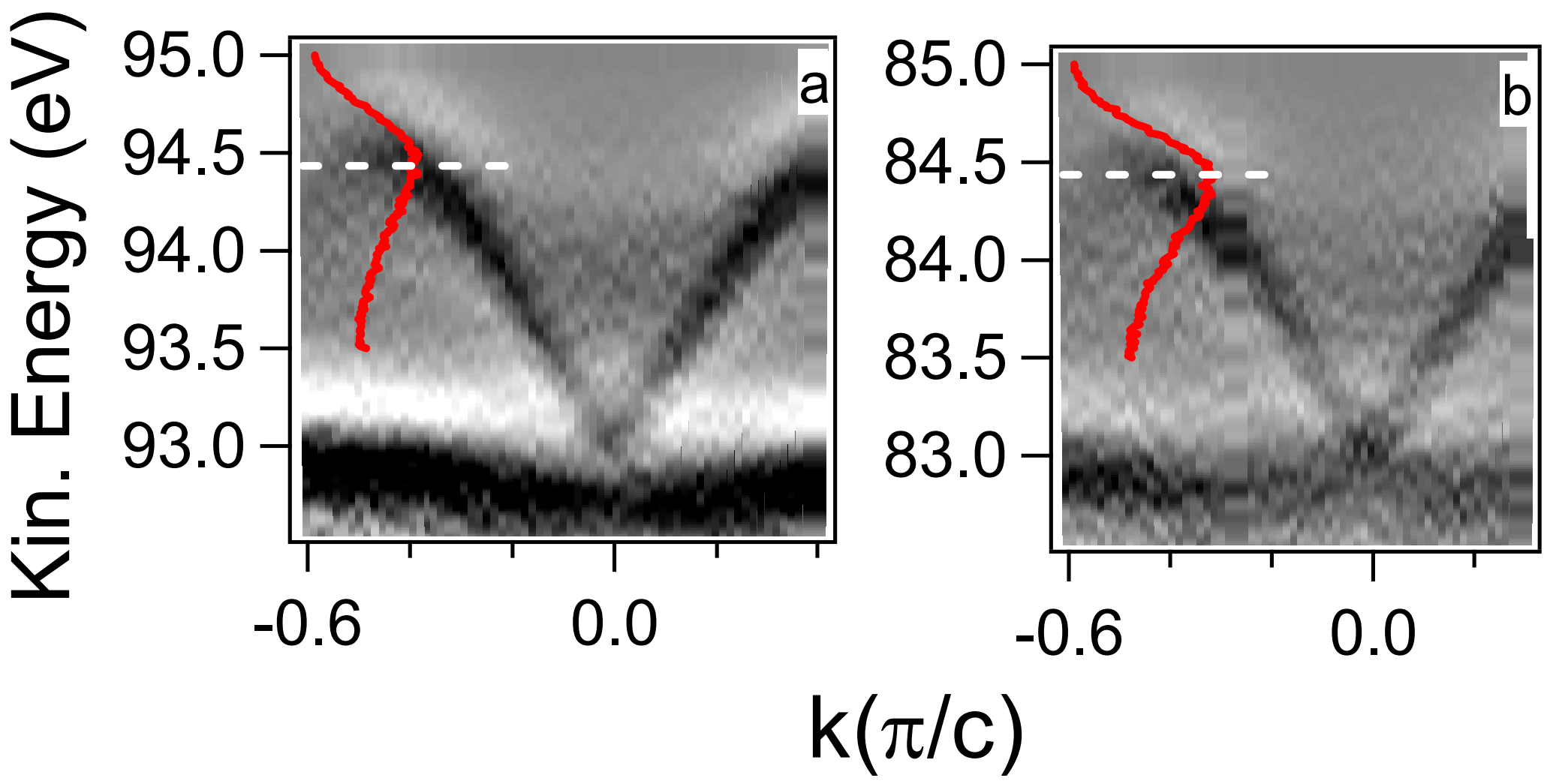

Fig. 2

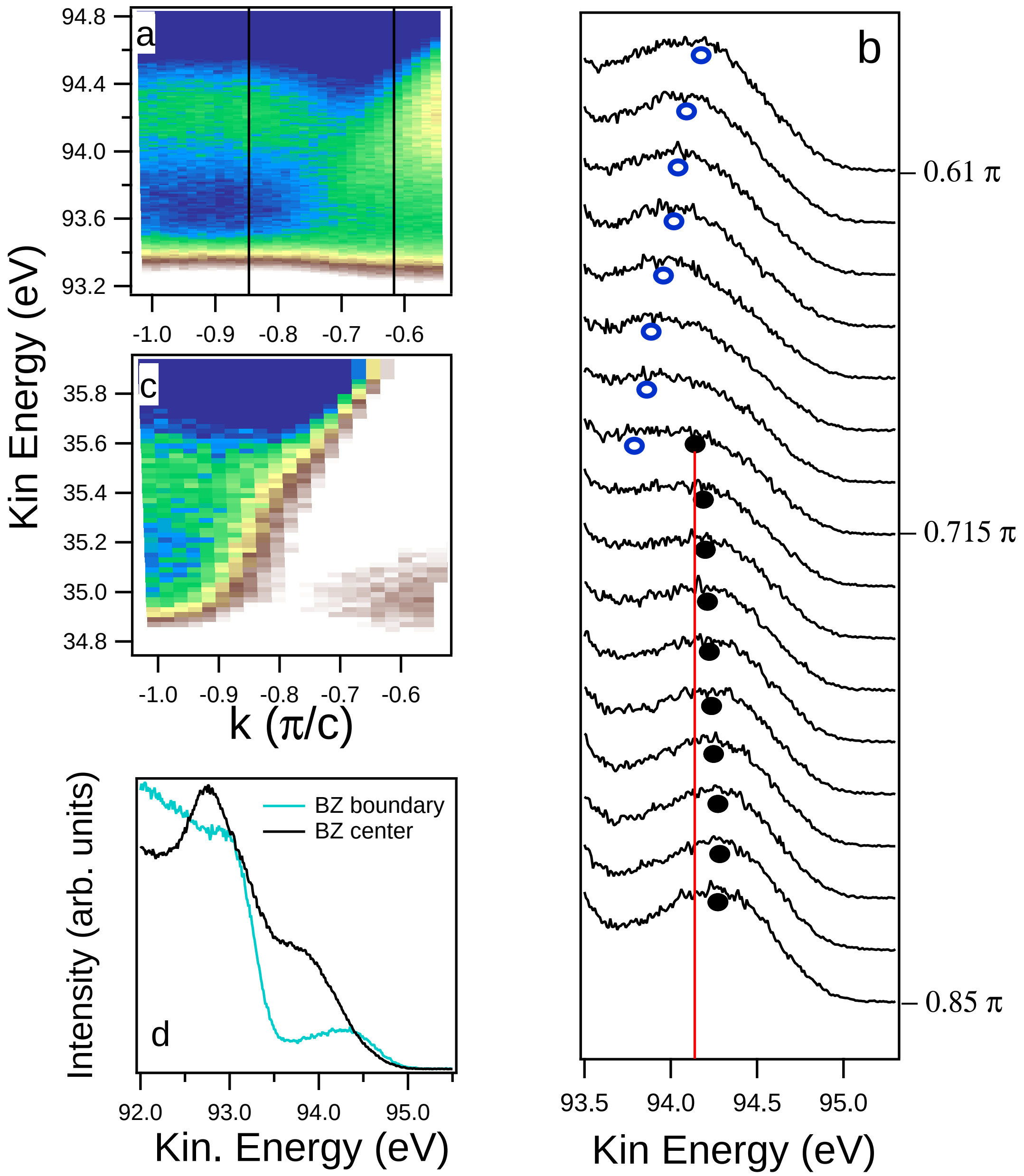

Fig. 3